\documentstyle[prd,aps]{revtex}

\begin{document}
\draft

\newcommand\lsim{\mathrel{\rlap{\lower4pt\hbox{\hskip1pt$\sim$}}
    \raise1pt\hbox{$<$}}}
\newcommand\gsim{\mathrel{\rlap{\lower4pt\hbox{\hskip1pt$\sim$}}
    \raise1pt\hbox{$>$}}}

\twocolumn[\hsize\textwidth\columnwidth\hsize\csname
@twocolumnfalse\endcsname
\preprint{PACS: 98.80.Cq\\
{}~hep-ph/yymmnn}

\title{Radiative seesaw and degenerate neutrinos}

\author{Borut Bajc$^{(1)}$ and Goran Senjanovi\'c$^{(2)}$}

\address{$^{(1)}${\it J. Stefan Institute, 1001 Ljubljana, Slovenia}}

\address{$^{(2)}${\it International Centre for Theoretical Physics,
34100 Trieste, Italy }}

\date{\today}
\maketitle

\begin{abstract} 
The radiative see-saw mechanism of Witten generates the right-handed 
neutrino masses in SO(10) with the spinorial $16_H$ Higgs field. 
We study here analytically the $2^{nd}$ and $3^{rd}$ generations 
for the minimal Yukawa structure containing $10_H$ 
and $120_H$ Higgs representations. In the approximation of small 
$2^{nd}$ generation masses and gauge loop domination we find the 
following results : (1) $b-\tau$ unification, (2) natural coexistence 
between large $\theta_l$ and small $\theta_q$, (3) degenerate neutrinos.
\end{abstract}


\vskip1pc]

{\it A. Introduction}. \hspace{0.5cm} SO(10) grandunified theory 
offers a natural and simple arena for the study 
of fermionic masses and mixings since it relates quark and lepton 
properties. Through the see-saw mechanism \cite{Minkowski:1977sc} 
and the resulting large mass for 
the right-handed neutrinos it explains the smallness of neutrino masses 
and offers a natural setting for leptogenesis \cite{Fukugita:1986hr}. 
It is tempting thus to 
construct the minimal realizations of the theory and see whether or not 
they can be realistic and predictive. 
The crucial point here is the minimal Yukawa structure especially if we 
stick to the renormalizable theory.  We know that the minimal Higgs $10_H$ 
does not suffice since the SM doublets in $10_H$ are SU(4)$_C$ singlets 
and thus would give the same quark and lepton structure; furthermore, with 
one Yukawa matrix no mixings are allowed. 

Two simple extensions are   possible: (i)  
$10_H+\overline{126}_H$; (ii) $10_H+120_H$.

(i) The first case is appealing since it simultaneously corrects bad mass 
relations $m_q=m_l$ through a Pati-Salam (2,2,15) field in 
$\overline{126}_H$ and provides a large mass for right-handed 
neutrinos through (1,3,10) in $\overline{126}_H$ 
\cite{Lazarides:1980nt,Babu:1992ia}. It has only $3+12=15$ real Yukawa 
couplings and can be considered the minimal supersymmetric GUT 
\cite{Aulakh:1982sw}. It was studied at length in the recent few years 
in the supersymmetric version \cite{Oda:1998na}, \cite{Fukuyama:2004xs}, 
where it appears to be realistic \cite{Bertolini:2005qb}. 
It gives $\theta_{13}\gsim 0.1$ and in the 
context of type II see-saw, large $\theta_{atm}$ is intimately 
related to the $b-\tau$ unification \cite{Bajc:2002iw}. 
Furthermore, in SO(10), as in any theory with gauged $B-L$, 
$R$-parity is a gauge symmetry \cite{Mohapatra:1986su} 
and it can be shown in this case to be an exact symmetry at low 
energies \cite{Aulakh:1997ba}, leading to the stable LSP, a natural 
candidate for the dark matter. 

(ii) This version has only $9$ real Yukawa 
couplings (see below). The charged fermions 
were studied in \cite{Matsuda:2000zp}, but the crucial point is the 
connection between neutrino and charged fermion masses that we address 
here. The seesaw mechanism takes the radiative form 
\cite{Witten:1979nr}: right-handed neutrino masses are generated at the 
two loop level utilizing a $16_H$ Higgs with $\langle 16_H\rangle\approx 
M_{GUT}\approx 10^{16}$ GeV. This proposal fell from grace due to the 
advent of low-energy SUSY, which inhibits radiative corrections to the 
superpotential. It can of course be implemented in ordinary SO(10), but 
there typically $M_{\nu_R}$ ends up being too small. Schematically,

\begin{equation}
M_{\nu_R}\approx \left({\alpha\over\pi}\right)^2Y_{10}
{M_R^2\tilde{m}\over M_{GUT}^2}\;,
\end{equation}

\noindent where $M_R$ is the scale of the breaking of SU(2)$_R$ symmetry 
and $\tilde{m}$ the effective susy breaking scale in the visible sector 
(in ordinary, nonsupersymmetric theories, the formula works with 
$\tilde{m}=M_{GUT}$). With low-energy supersymmetry, $\tilde{m}\approx 1$ 
TeV, this obviously fails, while without supersymmetry gauge coupling 
unification forces $M_R$ to lie much below $M_{GUT}$, which again fails 
unless some extra fine-tuning is done. On the other hand, as we argued in 
\cite{Bajc:2004hr}, this works nicely in split susy 
\cite{Arkani-Hamed:2004fb}.

This is the scenario we follow here. We focus on the minimal Yukawa 
structure 

\begin{equation}
W_Y=16_F^T\left(Y_{10}10_H+Y_{120}120_H\right)16_F
\end{equation}

\noindent
with $Y_{10}=Y_{10}^T$ and $Y_{120}=-Y_{120}^T$. We can diagonalize 
$Y_{10}$, thus we have $3$ real parameters, which together with the 
$3$ complex parameters of $Y_{120}$ add to $9$ in total. 
Still, the full 3 
generation case is rather involved and messy, needing numerical studies. 
In order to get some physical  insight and simple 
analytical results, we focus here on the heaviest two generations. 
In the limit of small $m_s/m_b$, $m_\mu/m_\tau$ 
and $m_c/m_t$ we find

(a) $b-\tau$ unification of $10_H$ remains valid;

(b) naturally small quark mixing angle $\theta_{bc}$;

(c) large $\theta_{atm}$ implies degenerate neutrinos.

In other words, up to corrections $m_2/m_3$, $b-\tau$ unification is still 
a prediction of the theory, in spite of the fact that the $120_H$ adds a 
(2,2,15) field. The prediction of degenerate neutrinos 
is remarkable, since this theory has only a 
canonical, type I, see-saw and it is often argued that degenerate 
neutrinos are in contradiction with type I see-saw. 

In arriving at the above results 
two assumptions were made: 1) the renormalizable 
interactions provide the complete picture, which amounts to 
neglecting all possible higher-dimensional operators (for an approach 
with higher dimensional operators see e.g. \cite{Dermisek:2005ij} 
and references therein); 2) as 
in the original work of Witten, the radiative seesaw is dominated 
by the gauge loop effects. Furthermore, we allow for no singlets 
or textures (for an opposite approach see e.g. \cite{ilja} and 
references therein).

\vspace{0.2cm}

{\it B. Charged fermions and Dirac neutrinos}. \hspace{0.5cm}
With this strategy we can write for the fermionic mass matrices

\begin{eqnarray}
M_D=M_0+M_2&\;\;,\;\;&M_U=c_0M_0+c_2M_2\;\;,\nonumber\\
M_E=M_0+c_3M_2&\;\;,\;\;&M_{\nu_D}=c_0M_0+c_4M_2\;\;,
\end{eqnarray}

\noindent
where

\begin{eqnarray}
M_0&=&Y_{10}\langle (2,2,1)_{10}^d\rangle\;,\\
M_2&=&Y_{120}\left(\langle (2,2,1)_{120}^d\rangle+
\langle (2,2,15)_{120}^d\rangle\right)\;,\\
c_0M_0&=&Y_{10}\langle (2,2,1)_{10}^u\rangle\;,\\
c_2M_2&=&Y_{120}\left(\langle (2,2,1)_{120}^u\rangle+
\langle (2,2,15)_{120}^u\rangle\right)\;,
\end{eqnarray}

\noindent
and

\begin{eqnarray}
c_3={\langle (2,2,1)_{120}^d\rangle-3
\langle (2,2,15)_{120}^d\rangle\over 
\langle (2,2,1)_{120}^d\rangle+
\langle (2,2,15)_{120}^d\rangle}\;,\\
c_4={\langle (2,2,1)_{120}^u\rangle-3
\langle (2,2,15)_{120}^u\rangle\over 
\langle (2,2,1)_{120}^d\rangle+
\langle (2,2,15)_{120}^d\rangle}\;.
\end{eqnarray}

We diagonalize $M_0$ ($Y_{10}$), which preserves the antisymmetry 
of $M_2$, and thus

\begin{eqnarray}
\label{m0}
M_0=\pmatrix{
  a
& 0
\cr
  0
& b
\cr}\;,\;
M_2=\pmatrix{
  0
& i\alpha
\cr
  -i\alpha
& 0
\cr}\;\;.
\end{eqnarray}

Notice that $a,b$ can be chosen to be positive real numbers. 
Similarly, $c_0$ can be taken real, whereas the other 
parameters, $c_{2,3,4}$ and $\alpha$ are in general complex numbers. 
In the limit $m_s=0$ one gets 

\begin{equation}
\label{abjealpha2}
ab=\alpha^2\;,
\end{equation}

\noindent
which shows that $\alpha$ must be real in this approximation. 
Similarly, $m_\mu=0$ ($m_c=0$) implies $c_3=\pm 1$ ($c_2^2=c_0^2$). 
In other words,

\begin{eqnarray}
\label{mdu}
M_D=\pmatrix{
  a
& i\alpha
\cr
  -i\alpha
& b
\cr}
\;,\;
M_U=c_0\pmatrix{
  a
& \pm i\alpha
\cr
  \mp i\alpha
& b
\cr}\;\;.
\end{eqnarray}

Obviously $\theta_U=\pm\theta_D$, while from 

\begin{eqnarray}
\label{menud}
M_E=\pmatrix{
  a
& \pm i\alpha
\cr
  \mp i\alpha
& b
\cr}\;,\;
M_{\nu_D}=\pmatrix{
  c_0a
& ic_4\alpha
\cr
  -ic_4\alpha
& c_0b
\cr}\;\;.
\end{eqnarray}

\noindent
one has $\theta_E=\pm \theta_D$, whereas $\theta_{\nu_D}$ is at this point 
undetermined. 

The quark mixing angle $\theta_q=\theta_U-\theta_D$ can thus take 
the following values:
$\theta_q=0\;\;{\rm or} \pm 2\theta_E\;\;$.

The situation in the leptonic sector is of course more complex, 
especially since we utilize the radiative seesaw mechanism. 

In the approximation of vanishing second generation 
masses the matrices $M_D$ and $M_E$ in (\ref{mdu}) and (\ref{menud}) 
are Hermitian, and thus our first important prediction follows:

\begin{equation}
m_b=m_\tau=a+b\;.
\end{equation}

The $b-\tau$ unification associate with $10_H$ Higgs field continues 
to be valid. For $c_3=1$ this is expected, since 
$\langle(2,2,15)_{120}^d\rangle\approx 0$, so that 
the SU(4) colour Pati-Salam (quark-lepton) symmetry would not be 
broken in the down quark and charged lepton sector. Surprisingly 
$c_3=-1$ also works, in spite of 
$\langle(2,2,15)_{120}^d\rangle\approx\langle(2,2,1)_{120}^d\rangle$. 

Before we proceed to discuss this at length, an important 
question can be posed: do we really need the seesaw mechanism? 
Could neutrinos be Dirac particles? It is often imagined that in 
this case the leptonic mixing angles could not be so different from 
the quark ones. Notice that here 
this would not be true. First of all, since $\nu_2$ 
and $\nu_3$ are not very hierarchical, $c_4/c_2$ is not fixed and 
$\theta_{\nu_D}$ is arbitrary. If they were 
very hierarchical, then clearly $\theta_{\nu_D}=\pm\theta_E$ as in the 
quark case. Still there is a completely consistent solution 
$\theta_l=2\theta_E$, which could obviously be 
any number and even very large. After all, $\theta_E$ between 
$20$ and $25$ degrees is definitely a natural value, or at least 
as natural as any other. 

What goes wrong of course are the values of neutrino masses; they are 
of the same order of magnitude of the charged fermions. This is why the 
seesaw mechanism is 
a must in a well defined, predictive, SO(10) theory. On the other 
hand the seesaw mechanism without grandunification to set the scale of 
the righthanded neutrino masses is of little use in the quantitative 
determination of light neutrino masses. 

\vspace{0.2cm}
{\it C. Not to forget the seesaw}. \hspace{0.5cm} 
Witten's two loop diagram \cite{Witten:1979nr} 
is proportional to the Yukawa $Y_{10}$, i.e. 

\begin{equation}
\label{mnr}
M_{\nu_R}\propto M_0\;.
\end{equation}

In turn, the light neutrino mass matrix $M_N$ takes the 
form 

\begin{equation}
M_N=M_{\nu_D}^TM_{\nu_R}^{-1}M_{\nu_D}\propto
c_0^2M_0-c_4^2M_2M_0^{-1}M_2\;.
\end{equation}

After some elementary algebra 

\begin{equation}
\label{mn}
M_N\propto\left(c_0^2-c_4^2\right)M_0\;.
\end{equation}

Since we are working in the basis of $M_0$ being diagonal, 
the first immediate consequence is that $\theta_N=0$, and 
thus the weak current leptonic mixing angle 
$\theta_l=\theta_{atm}=\theta_E$. From $\theta_{atm}\approx\pi /4$, 
only the solution $\theta_{bc}=\theta_q=0$ found before is the physical 
one (the other one $\theta_{bc}\approx\pi /2$ does not work). This 
cancellation between $\theta_U$ and $\theta_D$ is similar in spirit 
to \cite{Dorsner:2004qb}. In other 
words, the small quark and the large leptonic mixing angle can coexist 
naturally as the result of the spontaneous symmetry breaking of 
the quark-lepton symmetry valid at $M_{GUT}$. 

Even more interesting is the fact that eq. (\ref{mn}) gives 
degenerate neutrinos. Namely, $\theta_E\approx\pi /4$ implies 
necessarily $a\approx b$ in view of (\ref{abjealpha2}), or in other 
words, $M_0\propto I$. 

One can ask the question as why we chose the type I see-saw as the 
only source of neutrino masses. As is well known, in SO(10) one 
generically obtains also a type II see-saw generated through the 
vacuum expectation value of the SU(2)$_L$ triplet 
\cite{Magg:1980ut,Lazarides:1980nt,Mohapatra:1980yp}. 
In this case the triplet is the effective operator made of the product 
of two doublets in $16_H$. However it will be suppressed by the same 
two loop effect that enhances the type I through the 
suppression of the right-handed neutrino masses. In short, the ratio 
of type II over type I contributions is suppressed by $(\alpha/\pi)^4$, 
which implies that type II can be safely neglected.

\vspace{0.2cm}

{\it D. Discussion and outlook}. \hspace{0.5cm}
There are a number of important issues that ought to be addressed in 
order to have a full realistic theory of three generations, analogous 
to what has been achieved in the $10 + \overline{126}$  Yukawa case. 
We go through some central ones.

\begin{description}
\item[(i)] What about a radiative $\nu_{R}$ mass in the presence of a 
$120_{H}$ field? It can affect the simple, predictive result of Witten 
only if $120_{H}$ mixes through a $45_{H}$, which happens through a 
superpotential coupling 
$$\lambda 10_H 45_H 120_H\;.$$
If $\lambda \sim g $, 
the gauge coupling, there will be new diagrams exchanging both 
$10_H$ and $120_H$ which would spoil the prediction $\theta_N \sim 0$. 
Of course, a large $\theta_{\rm atm}$ solution remains possible; if 
anything, there would be more freedom and more chance that the full 
3 generation case works out. 

A small $\lambda$ is rather appealing and 
can be made natural by emposing a discrete symmetry 
$$(16,\overline{16})\to i(16,\overline{16})\;,\;
(10,120,45)\to -(10,120,45)$$
outside of $SO(10)$. This would ensure the stability of small 
$\lambda$ even in a strongly broken (split) supersymmetry or in 
a non-susy theory.

Furthermore, small $\lambda$ leads to the formation of domain walls 
generated by a nonvanishing $\langle 45_H\rangle \simeq M_{GUT}$. The 
subsequent evolution (disappearance) of these unstable domain walls has 
a remarkable consequence: they sweep away the magnetic monopoles 
\cite{Dvali:1997sa}. This 
provides the simplest and most elegant field theoretic 
explanation of the small monopole 
density of the universe (for a cosmological explanation see e.g. 
\cite{Stojkovic:2004hz}), a 
solution which needs no new fields or 
interactions and works independently of when inflation took place or 
how big the reheating temperature was. It is especialy appealing in view 
of the fact that the other simple solution \cite{Dvali:1995cj}, 
based on the non-restoration of symmetries \cite{Weinberg:1974hy} 
or the $U(1)_{em}$ breaking 
\cite{Langacker:1980kd} at high T faces trouble in gauge 
theories \cite{Bimonte:1995xs}, especially the supersymmetric ones 
\cite{Haber:1982nb}.

\item[(ii)] Which theory can give 
gauge coupling and $b-\tau$ unification consistent with symmetry 
breaking and phenomenology? The simplest solution, split 
supersymmetry with superheavy sfermions, works nicely for neutrino 
masses and gauge coupling unification, and $b-\tau$ unification 
favours small $\tan{\beta}$ \cite{Giudice:2004tc}. 
In strongly split supersymmetry there is an issue of stable 
gluinos though \cite{Arvanitaki:2005fa}. 
Notice however that for $c_0^2 \simeq c_4^2$ neutrino masses get 
further suppressed, beyond the generic see-saw. 
This means that sfermions may lie below the 
unification scale, i.e. $\tilde{m}\ll M_{GUT}$, which somewhat 
alleviates the above problem.

Another possibility is ordinary nonsupersymmetric SO(10), but 
again some extra fine-tuning would be mandatory. 
It has to be noticed however, that in order to continue having only 
two Yukawa matrices, a Peccei-Quinn-type symmetry must be imposed 
in this case. 

\item[(iii)] What about the impact of running from $M_{GUT}$ 
down to $M_Z$? 
In this analytic approach we only wanted to get a qualitative insight 
into the pattern of neutrino masses and mixings. 
One should definitely run, though, when a 3 generation 
numerical study is built up around our solution. 

A serious challenge appears to be the prediction of all three neutrinos 
being degenerate. The full three generations numerical study can 
invalidate this, since eq. (\ref{mnr}) does not follow automatically 
even with the assumption of gauge loop domination. 
A way out in such a case could be allowing for larger $\lambda$.

\item[(iv)] Is it really necessary to neglect the higher dimensional 
operators? After all, the second generation masses remain small, 
compatible with zero, even in this case. The only problem is 
represented by a possible direct contribution to the right-handed 
neutrino mass squared of the order $M_{GUT}^2/M_{Planck}$. 
This is of the same order of magnitude as the radiatively generated 
term. However, there is no need for it to be proportional to 
$Y_{10}$. It is for this reason that we need to assume that at least 
this operator is suppressed. 

\end{description}
 
In summary, a radiative see-saw mechanism offers an alternative 
simple SO(10) theory if one sticks to the minimal Yukawa structure 
with $10_H$ and $120_H$. 

\vskip 0.3cm

We thank K. Babu, J. Bagger, I. Gogoladze, A. Melfo, 
M. Nemev\v sek, A. Romanino and F. Vissani for valuable 
comments and discussions. 
The work of G.S. was supported by EEC (TMR contracts ERBFMRX-CT960090 
and HPRN-CT-2000-00152), the work of B.B. by the Ministry of Education, 
Science and Sport of the Republic of Slovenia.

 \end{document}